\documentclass[fleqn,usenatbib]{mnras}
\usepackage{newtxtext,newtxmath}
\usepackage[T1]{fontenc}

\usepackage[figuresleft]{rotating}
\usepackage{graphicx}
\usepackage{multirow}
\usepackage{lipsum}
\usepackage{blindtext}
\usepackage{amsmath}
\usepackage{xspace}
\usepackage{pdflscape}
\usepackage{afterpage}
\usepackage{gensymb}
\usepackage{booktabs}
\usepackage{bold-extra}
\usepackage[usenames, dvipsnames]{color}
\usepackage{adjustbox}

\title[SHAM in IllustrisTNG100]{SHAM through the lens of a hydrodynamical simulation}
\author[Favole et al. 2021]
{\parbox[t]{\textwidth}{\vspace{-0.8cm}Ginevra Favole,$^{1}$\thanks{E-mail: ginevra.favole@epfl.ch} Antonio D. Montero-Dorta,$^{2,3}$ M. Celeste Artale,$^{4}$ Sergio Contreras,$^5$ Idit Zehavi,$^6$ Xiaoju Xu$^6$}
\vspace*{15pt}\\
$^1$Institute of Physics, Laboratory of Astrophysics, Ecole Polytechnique F\'ed\'erale de Lausanne (EPFL), Observatoire de Sauverny, 1290 Versoix, Switzerland.\\
$^{2}$ Departamento de F\'isica, Universidad T\'ecnica Federico Santa Mar\'ia, Casilla 110-V, Avda. Espa\~na 1680, Valpara\'iso, Chile\\
$^3$Instituto de F\'isica, Universidade de S\~ao Paulo, Rua do Mat\~ao 1371, CEP 05508-090, S\~ao Paulo, Brazil \\
$^4$Institut f\"{u}r Astro- und Teilchenphysik, Universit\"{a}t Innsbruck, Technikerstrasse 25/8, 6020 Innsbruck, Austria.\\
$^5$Donostia International Physics Center (DIPC), Manuel Lardizabal Ibilbidea, 4, 20018 Donostia, Gipuzkoa, Spain.\\
$^6$Department of Physics, Case Western Reserve University, Cleveland, OH 44106, USA.
\vspace{-1.2cm}}
\date{}
\begin{document}
\pubyear{2020}
\maketitle
\begin{abstract}
We use the IllustrisTNG100 hydrodynamical simulation to study the dependence of the galaxy two-point correlation function on a broad range of secondary subhalo and galactic properties. We construct galaxy mock catalogues adopting a standard sub-halo abundance matching scheme coupled with a secondary assignment between galaxy colour or specific star formation rate and the following subhalo properties: starvation redshift z$_{\rm starve}$, concentration at infall, overdensity $\delta_R^{\rm env}$, tidal anisotropy $\alpha_R$, and tidal overdensity $\delta_R$. The last two quantities allow us to fully characterise the tidal field of our subhaloes, acting as mediators between their internal and large-scale properties. The resulting mock catalogues overall return good agreement with the IllustrisTNG100 measurements. The accuracy of each model strongly depends on the correlation between the secondary galaxy  and subhalo properties employed. Among all the subhalo proxies tested, we find that z$_{\rm starve}$ and $c_{\rm infall}$ are the ones that best trace the large-scale structure, producing robust clustering predictions for different samples of red/blue and quenched/star-forming galaxies.
\end{abstract}


\section{Introduction}
\label{sec:intro}

Connecting the properties of galaxies to those of dark-matter haloes is 
a fundamental step in the extraction of cosmological information from measurements 
of galaxy clustering \citep[][]{Wechsler2018} along with a necessary validation for theories of 
galaxy formation inside haloes. It is today accepted that the mass of the 
hosting halo correlates with the stellar mass of the galaxy located at its
centre \citep{Behroozi2010,Guo2010,Moster2013,Matthee2017}, with its size \citep{Rodriguez2020} and with the total 
galaxy content or {\it{occupation}} of the halo (e.g., \citealt{Berlind2002, Zehavi2005,Artale2018,Bose2019,Hadzhiyska2020B,Xu2021}), to 
name but a few. These and other correlations are nothing but a measurable manifestation 
of the multiple physical processes that take place inside haloes, which shape 
the evolution of its baryonic content. 

One of the main methods to perform the aforementioned connection is the 
{\it{sub-halo abundance matching technique}} \citep[SHAM;][]{Conroy2006,Behroozi2010,Trujillo-Gomez2011,Favole2016,Chaves-Montero2016,Favole2017,Rodriguez-Torres2016,Rodriguez-Torres2017,Guo2016,Guo2016b,Contreras2020,Contreras2021,Hadzhiyska2021}. In SHAM, galaxies from 
an observational (or synthetic) data set are linked to haloes from an N-body numerical 
simulation by matching their number densities, assuming a one-to-one correspondence
between {\it{primary}} halo and galaxy properties. For haloes, either a halo mass 
or a velocity-related quantity (in their multiple forms) are assumed, on the basis that 
they are the main determinants of halo clustering. Analogously, either stellar mass
or luminosity is chosen for the galaxy set, since they are easy to measure and 
are the properties that are known to correlate better with halo mass/velocity. To this simple 
prescription, a parametrized scatter in the halo-galaxy correspondence is added in order to 
account for the stochasticity in the way that galaxies populate haloes. 
The SHAM technique has been employed successfully in a number of clustering 
works accross different redshifts and for several galaxy populations \citep{Favole2016,Favole2017,Rodriguez-Torres2016,Rodriguez-Torres2017,Guo2016,Granett2019,Jullo2019}. 

As mentioned above, the majority of the SHAM modelling is performed on the basis of a single 
halo/galaxy property. However, {\it{at fixed halo mass}}, the clustering of haloes 
is known to depend on secondary halo properties such as formation redshift (which encodes the assembly history of haloes), concentration, and spin (see, e.g., \citealt{Gao2005, Wechsler2006,Angulo2008,Dalal2008,Salcedo2018,SatoPolito2019, Johnson2019, Mansfield2020, Tucci2020}). This effect, broadly referred to as {\it{secondary halo bias}}, is expected to have a manifestation on the galaxy population (see discussion in, e.g., \citealt{Miyatake2016, More2016,Zhu2006,MonteroDorta2017, Niemiec2018,Zehavi2018,MonteroDorta2020B, Obuljen2020, Salcedo2020}), which has forced halo--galaxy connection models to adjust. 

Additional dependencies of galaxy and halo clustering have already been implemented into the SHAM formalism. Namely, in \citet{Hearin2013}, the SHAM modelling of the differential clustering of red and blue galaxies is performed by including a secondary dependence on the ``starvation redshift", related to the formation redshift of haloes. This incarnation of SHAM is called {\it{age matching}}, in reference to the fact that an additional matching {\it{at fixed halo mass}} (or maximum rotational velocity $V_{\rm max}$) is performed based on a halo-age-related quantity. 

In the context of secondary halo bias, several works have attempted to provide an explanation for the physical mechanisms behind this convoluted set of trends \citep[e.g.,][]{Dalal2008,Hahn2009,Borzyszkowski2017,Musso2018,Paranjape2018, Mansfield2020,Paranjape2020,Ramakrishnan2019,Ramakrishnan2020,Zjupa2020, Tucci2020}. As a result of these efforts, a common picture is starting to emerge, where {\it{halo assembly bias}} (the secondary dependence on halo accretion history) might be intimately connected to environmental processes that take place in the cosmic web. In essence, assembly bias might be the result of the truncation of accretion history in a population of smaller mass haloes, which could be more likely in certain environments (i.e. filaments) than others (nodes), see an illustrative description of these processes in \citet{Borzyszkowski2017} and \citet{Musso2018}.

The cosmic web environment can be characterised in multiple ways. One simple option is to measure the halo-centric density in spheres of a certain radius (i.e. $\sim$ 5 $h^{-1}$Mpc). Also informative (and in a sense complementary) is the tidal anisotropy parameter $\alpha_R$, which can be determined from the eingenvalues of the tidal tensor. The tidal tensor describes the gravitational effect exerted by the global distribution of matter around a point, so its anisotropy allows us to characterise the cosmic web: large values of $\alpha_R$ correspond to filaments and sheets, whereas smaller values are associated with regions where matter is accreted from all directions, i.e., nodes. Importantly, in a series of works \citep{Paranjape2018, Paranjape2020,Ramakrishnan2019,Ramakrishnan2020}, it has been recently claimed that assembly bias correlates directly with $\alpha_R$, which is likely a reflection of the environment-related truncation of accretion mentioned above \citep{Hahn2009,Borzyszkowski2017,Musso2018}.

Hydrodynamical simulations are laboratories for galaxy-formation physics which are ideal to test halo-galaxy linking techniques, since both the dark-matter and the baryonic components of haloes in detail. In this context, the recently released IllustrisTNG \citep{Pillepich2018,Nelson2019}
suite of hydrodynamical simulations offers some advantages, such as the large size of some of their boxes (up to a side length of 205 $h^{-1}$Mpc). IllustrisTNG has already shed light onto crucial aspects of the connection between galaxies and haloes. \cite{MonteroDorta2020B} showed how secondary halo bias would manifest itself in the clustering of the central galaxy population when this is selected on the basis of several galaxy properties. \citet{Bose2019} and \citet{Hadzhiyska2020} addressed halo occupation in IllustrisTNG, demonstrating that the basic (mass-based) halo occupation distribution (HOD) ansatz underpredicts the real-space correlation function in the largest IllustrisTNG box. They also discuss several ways of ``augmenting" the modelling by including secondary halo dependencies, a work that was subsequently extended in \cite{Hadzhiyska2020B}. \cite{Contreras2020} used IllustrisTNG to test a novel and flexible modification of SHAM where an arbitrary amount of assembly bias can be incorporated into the modelling.

The main goal of this paper is to build on previous efforts and revisit the SHAM method using IllustrisTNG. The philosophy behind our approach follows that of \cite{Hearin13}, in that we are interested in modelling galaxy populations selected by colour and star-formation rate by including both secondary dependencies on halo and galaxy properties. We also test the inclusion of new physically motivated halo properties within the formalism, including the anisotropy tidal parameter $\alpha_R$ and the density of the environment around haloes. 

The paper is organised as follows. Section~\ref{sec:data} provides a brief description of the TNG100 simulation data analysed in this work, including halo/galaxy properties (\S~\ref{sec:properties}) and sample selection (\S~\ref{sec:selection}). The environmental properties (dark matter density contrast and tidal-field measurements) are described in \S~\ref{sec:environment}. The methodology used to measure galaxy clustering in TNG100 and to estimate the associated uncertainties is detailed in Section~\ref{sec:clustering_measurements}. Our SHAM implementation is described in \S~\ref{sec:multiSHAM}. Section~\ref{sec:results} presents the main results of our analysis: the correlations between secondary halo and galaxy properties (\S~\ref{sec:secpropcorrel}) and the clustering outcomes~\ref{sec:clustering_results}. In Section~\ref{sec:discussion} we summarize and discuss our findings. 

The IllustrisTNG simulations adopt the standard $\Lambda$CDM cosmology  \citep{Planck2016}, with parameters $\Omega_{\rm m} = 0.3089$,  $\Omega_{\rm b} = 0.0486$, $\Omega_\Lambda = 0.6911$, $H_0 = 100~\,h\, {\rm km\, s^{-1}Mpc^{-1}}$ with $h=0.6774$, $\sigma_8 = 0.8159$ and $n_s = 0.9667$.

\section{Simulation data}
\label{sec:data}
In this paper we use the  magneto-hydrodynamical cosmological simulations IllustrisTNG \citep{Weinberger2017,Marinacci2018,Naiman2018,Nelson2018,Pillepich2018,Springel2018}. The IllustrisTNG represent a revised version of the Illustris simulations \citep{Vogelsberger2014, Vogelsberger2014b, Genel2014}, run with the {\sc arepo} moving-mesh code \citep{Springel2010}.
Among others, the updates on IllustrisTNG include magnetic fields, a revised scheme for galactic winds, and a new model for AGN feedback. The simulations include sub-grid models that account for star formation, chemical enrichment from SNII, SNIa, and AGB stars, stellar feedback, radiative metal-line cooling, and AGN feedback.

The IllustrisTNG suite includes different sizes and resolutions. In this work we made use of the IllustrisTNG100-1 (\textquoteleft\textquoteleft TNG100", hereafter) run, and its dark matter only counterpart IllustrisTNG100-1-DMO (\textquoteleft\textquoteleft TNG100-DMO", hereafter). These are the second largest simulated boxes and with the highest resolution available on the database\footnote{\url{http://www.tng-project.org}}. The TNG100 and TNG100-DMO represent a cubic box of side $L_{\rm box}=75\,h^{-1}$~Mpc, with periodic boundary conditions. 
The TNG100 run follows the dynamical evolution of initially  1820$^3$ gas cells of mass $9.4 \times 10^5 h^{-1} {\rm M_{\odot}}$, and dark-matter particles of mass $5.1 \times 10^6 h^{-1} {\rm M_{\odot}}$.
TNG100-DMO was run with $N_{\rm p}=1820^3$ dark matter particles of mass  $m_{\rm p}=6 \times 10^6 h^{-1} {\rm M_{\odot}}$.

The choice of TNG100 is motivated by its high resolution, as compared to the larger TNG300 box ($L_{\rm box}=205\,h^{-1}$~Mpc). We have in fact checked that although the TNG300 volume is beneficial in terms of the computation of clustering, some of the halo properties that we determine here can be severely affected by its lower resolution. 

The TNG and TNG-DMO simulations have been already employed to investigate several features on the galaxy--dark matter halo connection \citep{Bose2019,Contreras2020,Gu2020,Hadzhiyska2020,Hadzhiyska2020B,Shi2020,MonteroDorta2020B,MonteroDorta2020C} proving they are a suitable tool for the goal of this paper.
In the next section we describe the set of properties implemented.


\subsection{Halo and galaxy properties}
\label{sec:properties}
We use the galaxy and dark matter catalogues from TNG100 and TNG100-DMO available on the database. The dark-matter haloes (also referred as \textquoteleft \textquoteleft groups" in TNG) are identified with a friends-of-friends (FOF) algorithm using a linking length of 0.2 times the mean inter-particle separation \citep{Davis1985}, while the gravitationally bound substructures (also called \textquoteleft \textquoteleft subhaloes" in TNG) are identified using the SUBFIND algorithm \citep{Springel2001,Dolag2009}. The TNG subhaloes can be either central or satellite structures.

In order to develop the SHAM modeling, we combine the subhalo properties from the TNG100 hydrodynamical simulation with those belonging to its dark-matter-only counterpart, TNG100-DMO. 
In TNG100, subhaloes with non-zero stellar mass component are defined as galaxies. Each dark-matter halo can contain a central galaxy and several satellites, depending on the size and mass of the halo. We use the following properties from the subhaloes of the TNG100 simulation:
\begin{itemize}
    \item ${\rm M_\ast{}}$ [$h^{-1} {\rm M_{\odot}}$]: stellar mass computed as the total mass of the stellar particles bound to each subhalo.
    \item $(g-i)$: intrinsic galaxy colour from the IllustrisTNG database. We note that it does not include the attenuation produced by dust.
     \item SFR [${\rm M_{\odot}~yr^{-1}}$]: galaxy star formation rate. It is defined as the sum of the star formation rate of the gas cells in each subhalo.
    \item sSFR [${\rm yr^{-1}}$]: specific star formation rate, computed as sSFR = SFR/M$_\ast{}$.
\end{itemize}

From TNG100-DMO, the following properties are employed:
\begin{itemize}
\item $V_{\rm peak}$ [km\,s$^{-1}$]: maximum circular velocity estimated over the entire history of the subhalo.
  \item M$_{\rm vir}$ [$h^{-1} {\rm M_{\odot}}$]: virial mass computed as the total mass of dark matter particles\footnote{For the TNG100 simulation that includes baryons, the virial mass is computed accounting for all the total mass, i.e., dark matter, gas, and stars.} within a radius R$_{\rm vir}$, where the enclosed density equals 200 times the critical density.
  \item M$_{\rm subhalo}$ [$h^{-1} {\rm M_{\odot}}$]: subhalo mass, computed as the total number of dark-matter particles times the mass of each individual particle in the subhalo.
  \item z$_\text{char}$: characteristic redshift defined as the redshift at which the subhalo first reaches a mass of $10^{12}$ [$h^{-1} {\rm M_{\odot}}$]. For haloes that never attain this mass, z$_\text{char}=0$.
  \item z$_{\rm 1/2}$: formation redshift, computed as the redshift at which, for the first time, half of the halo mass at $z=0$ has been accreted into a single subhalo. 
  \item  z$_\text{acc}$: accretion redshift defined as the redshift after which a subhalo always remains a subhalo. For parent haloes z$_\text{acc}=0$.
  \item z$_\text{starve}$: starvation redshift defined as in \citet{Hearin2013}:
  \begin{equation}\label{eq:zstarve}
    z_{\rm starve} = {\rm Max}(z_{\rm 1/2},z_{\rm char},z_{\rm acc}).
  \end{equation}
  \item c$_{\rm infall}$: concentration at the time of infall, i.e.,  when a halo falls within the virial radius of a larger halo, thus becoming a subhalo. As shown by \citet{Bullock2001} and \citet{Gao2007}, a good proxy for concentration is provided by the ratio between the maximum and virial circular velocities of the halo. We therefore define:
   \begin{equation}
  c_{\rm infall} = \frac{V_{\rm max}}{V_{\rm vir}}.
  \end{equation} 
	The concentration at infall is preferable to the virial quantity as it is a better proxy for $V_{\rm peak}$, correctly tracing the halo-subhalo hierarchy.
	\item $N_{\rm p}^{\rm{(halo)}}$: number of particles per resolved halo. This corresponds to the \textquoteleft\textquoteleft SubhaloLen" TNG100-DMO property.
\end{itemize}


\subsection{Sample selection}
\label{sec:selection}
\begin{figure}
\centering
\includegraphics[width=\linewidth]{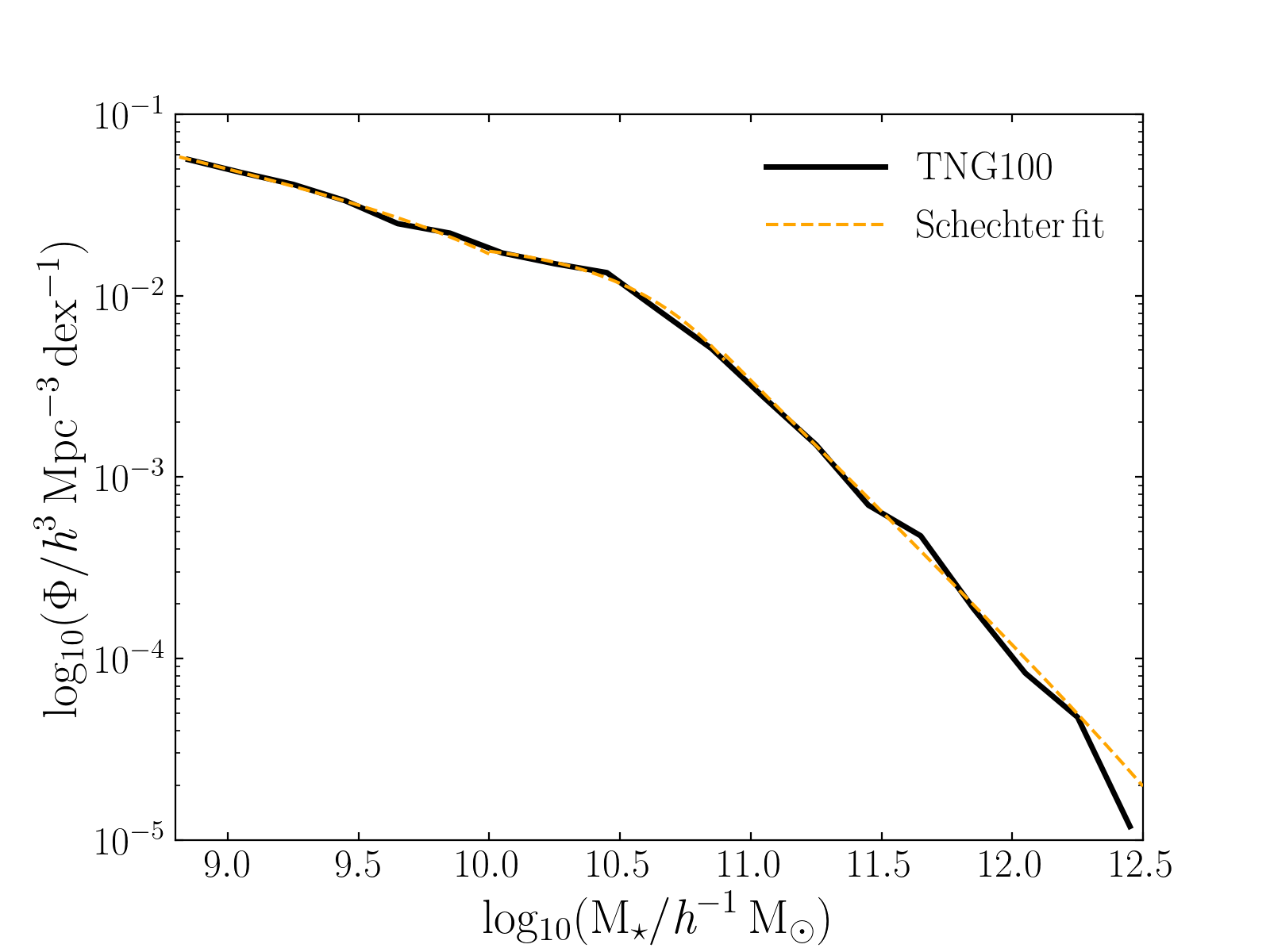}\\
\caption{TNG100 stellar mass function and corresponding Schechter fit given in Eq.\,\ref{eq:schechter}.}
\label{fig:smf}
\end{figure}
\begin{figure}
\centering            
\includegraphics[width=0.8\linewidth]{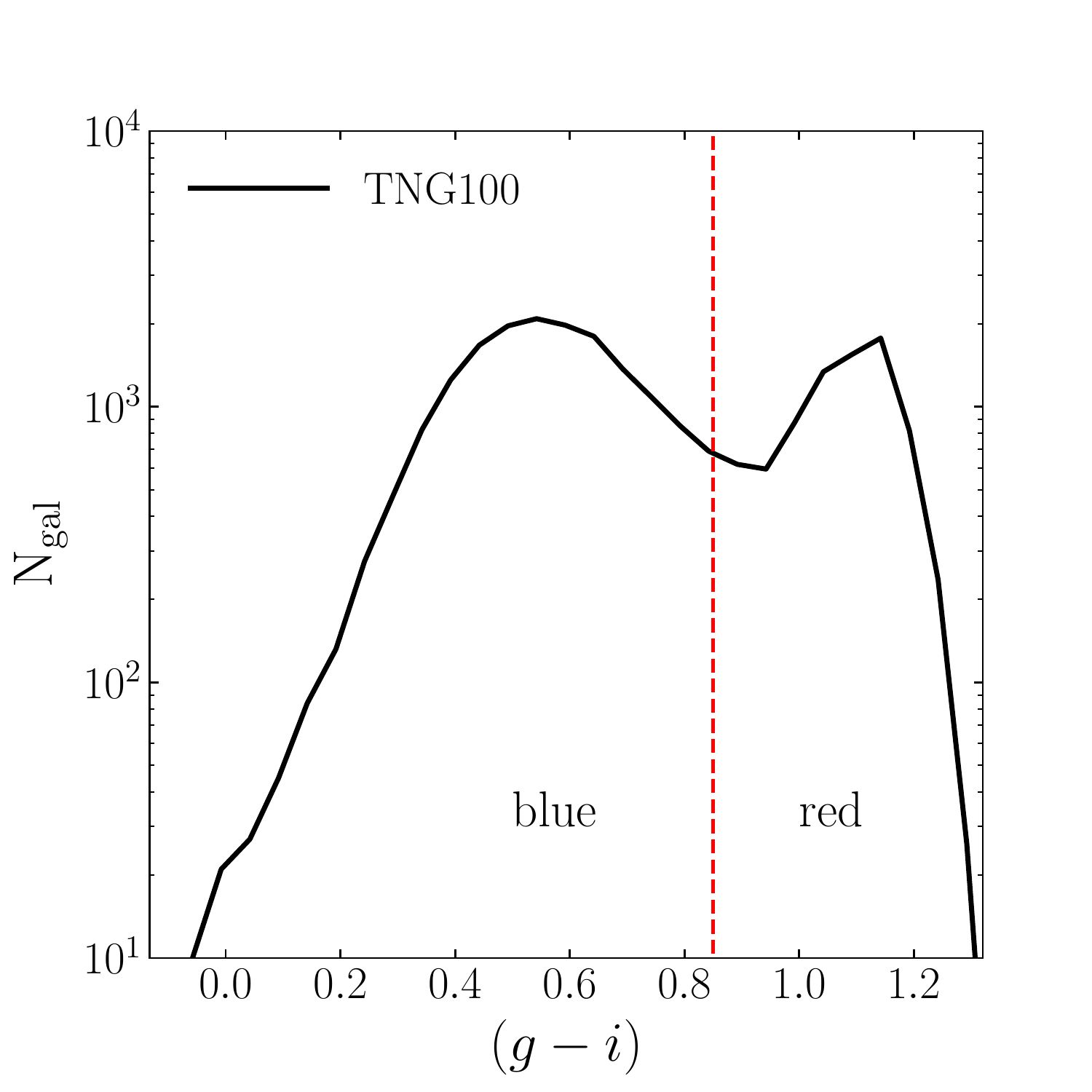}
    \includegraphics[width=0.8\linewidth]{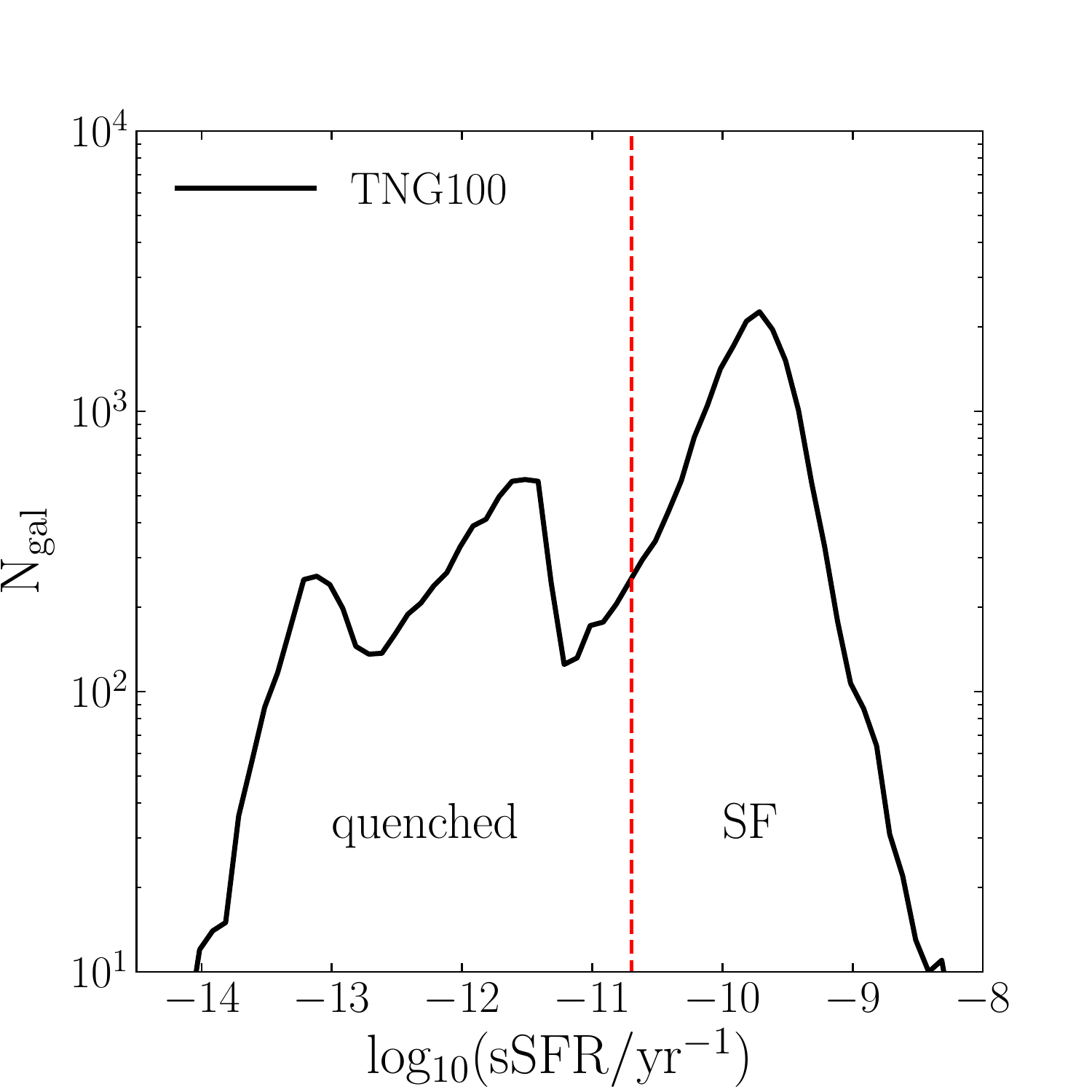}
        \caption{\textit{Top:} TNG100 colour distribution. The vertical line at $(g-i)=0.85$ denotes the cut that we apply to separate the red (right) from the blue (left) galaxy population. \textit{Bottom:} specific SFR distribution with the $\rm log(sSFR/yr^{-1})=-10.7$ cut which divides star-forming (right) from quenched (left) galaxies.} 
    \label{fig:dist}
\end{figure}
Our TNG100 galaxy and subhalo samples are selected by imposing two minimal cuts: ${\rm log (M_\star/{\textit h^{-1}}M_{\odot})}>8.75$, ${\rm log (M_{vir}/{\textit h^{-1}}M_{\odot})}>9.7$. These conditions eliminate the low-mass end of the subhalo (galaxy) distribution, which is not interesting for the current analysis, making the final catalogue more manageable. They also contribute to discarding a fraction of dark matter subhaloes which are below the resolution limit we adopt in Sec.\,\ref{sec:tidal} to calculate the halo tidal properties. We further remove from the resulting sample all subhaloes with non-physical values of the concentration at infall, i.e.,  (c$_{\rm infall}=0$ and c$_{\rm infall}=\infty$). These are spurious objects representing less than 0.2\% of the selection.

\begin{figure*}
    \centering\vspace{-0.4cm}\hspace{-0.5cm}
\includegraphics[width=1.15\linewidth]{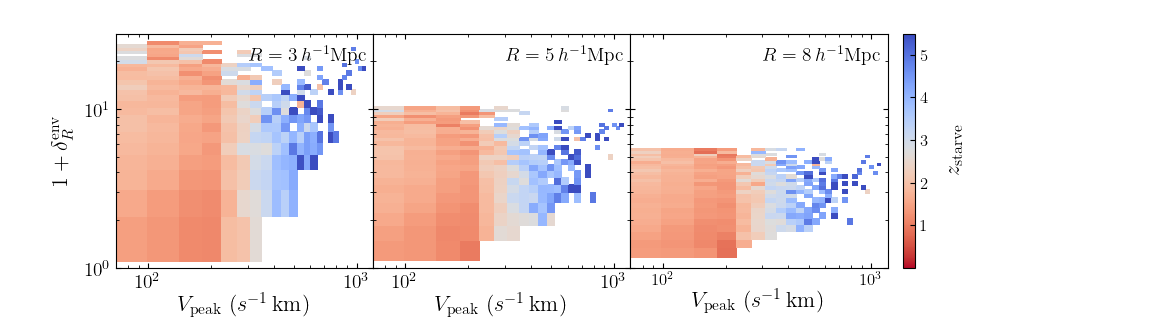}\vspace{-0.8cm}
\caption{Subhalo overdensity as a function of $V_{\rm peak}$, color-coded with z$_{\rm starve}$ for the subhaloes in TNG100-DMO. From left to right we show the results with a radius of 3, 5 and 8 $h^{-1}$Mpc. Here we show the average $z_{\rm starve}$ value in 30 bins of $V_{\rm peak}$ and subhalo overdensity.}
\label{fig:denscele}
\end{figure*}

Figure \ref{fig:smf} shows the stellar mass function of our final TNG100 galaxy sample obtained by applying the minimal cuts above. Overplot is a composite Schechter fit with form:
\begin{equation}
    \begin{aligned}
\Phi({\rm logM})={\rm ln}(10)\Phi^{\star}10^{(\alpha+1)({\rm log M}-{\rm logM^{\star}})}\\
\times \exp{[-10^{({\rm logM}-{\rm logM}^{\star})}]},
    \label{eq:schechter}
    \end{aligned}
\end{equation}
where ($\Phi^{\star}$, ${\rm logM}^{\star}$, $\alpha$) take the  values:
\begin{equation}
        \begin{aligned}
        &(7.288\times10^{-3}, 10.464, -1.333)\,\,\,\,\,{\rm at}\,\,\,\,{\rm log(M/h^{-1}M_{\odot})}<10.0,\\
        &(1.011\times10^{-2}, 10.673, -0.955)\,\,\,\,\,{\rm at}\,\,\,\,10.0\leq \log{\rm (M}/h^{-1}{\rm M_{\odot}})<11.1,\\
       &(1.158\times10^{-6}, 13.171, -2.433)\,\,\,\,\,{\rm at}\,\,\,\,\log{\rm(M}/h^{-1}{\rm M_{\odot}})\ge 11.1.
       \label{eq:schechter_params}
\end{aligned}
\end{equation}

As Figure \ref{fig:smf} shows, the Schechter fit provides a good analytical description of the TNG100 stellar mass function. This model will be useful in the context of the SHAM prescription implemented in this work. The fit shown here is not smooth as it is the sum of three Schechter functions with parameters given in Eq.\,\ref{eq:schechter_params}. This specific shape has been chosen to maximize the agreement with the TNG100 stellar mass function.

Another important element in our analysis is the distribution of galaxy colours in TNG100. The top panel in Figure \ref{fig:dist} shows the $(g-i)$ colour distribution of the TNG100 galaxy sample with the characteristic blue (left) and red (right) peaks. In order to study the galaxy clustering dependence on colour, in the analysis we separate the two populations by imposing a $(g-i)=0.85$ cut. Note that this cut is solely based on the TNG100 colour distribution of Figure \ref{fig:dist}.

About 23\% of the TNG100 galaxies are quenched with $\rm SFR=0$. This produces a singularity in the $\rm log(sSFR/yr^{-1})$ distribution, which hinders the analytical fitting and subsequent modelling shown in  Sec.\,\ref{sec:methodology}. To circumvent this problem, we randomly scatter the null SFRs using a Gaussian distribution with $\sigma=4\times 10^{-4}\,[{\rm{M_{\odot}\,yr^{-1}}}]$. This specific value has been chosen so that it returns a quenched peak in the $\rm log(sSFR/yr^{-1})$ distribution with no overlap with the star-forming one. The final sSFR distribution including the scatter is shown in the bottom panel of Figure \ref{fig:dist}. We investigate the clustering dependence on sSFR by separating the quenched from the star-forming population at $\rm log (sSFR/yr^{-1})=-10.7$. This value is similar with previous demarcations employed in the context of IllustrisTNG \citep[see e.g.,][]{Donnari2019}.


\subsection{Environmental properties}
\label{sec:environment}
One of the main goals of this work is to test the inclusion of additional dependencies of subhalo clustering into the SHAM ansatz. Motivated by recent secondary halo bias results (e.g., \citealt{Paranjape2018, Ramakrishnan2019}), we choose to focus on halo environment. In particular, we test different prescriptions of environment, both based on the subhalo occupancy number within a certain radius and on the halo virial mass. 
In this section, we describe how to measure such environmental properties, which have been computed by using the TNG100-DMO simulation.

\subsubsection{{\bf Subhalo overdensity}}
\label{sec:density_contrast}
A fundamental quantity used to characterise the environment of a subhalo is its overdensity $\delta_R^{\rm env}$. This is computed as the number density of subhaloes within a sphere of radius $R$, normalised by the total number density of subhaloes in the box \citep[e.g.,][]{Artale2018,Bose2019}. The subhalo overdensity is also a biased tracer of the DM density contrast.

We compute such density by adopting periodic boundary conditions for three different radii, namely, 3, 5, 8\,$h^{-1}$Mpc. This environmental property will be used as a secondary subhalo property in the analysis.

Figure\,\ref{fig:denscele} shows the distribution of the subhalo overdensity at three different radii as a function $V_{\rm peak}$, color-coded by z$_{\rm starve}$. The first thing to notice is that, while low-V$_{\rm peak}$ (typically low-mass) subhaloes live in all types of environments, there is a preference for higher-V$_{\rm peak}$ (typically higher-mass) subhaloes to inhabit slightly denser environments. This trend is progressively washed out as we increase the radius. Regarding the secondary dependence on z$_{\rm starve}$, Figure\,\ref{fig:denscele} seems to indicate that higher-V$_{\rm peak}$ subhaloes suffer the truncation of their accretion earlier (they are older), as compared to their lower-V$_{\rm peak}$ counterparts. At fixed V$_{\rm peak}$, Figure\,\ref{fig:denscele} does not display a visible correlation between z$_{\rm starve}$ and the subhalo overdensity. 

The top panel in Figure\,\ref{fig:ad} displays the spatial distribution of the subhaloes in a TNG100-DMO slice $10\,h^{-1}$Mpc thick. The color code represents the subhalo overdensity for a radius of 3\,$h^{-1}$Mpc. Subhaloes located in knots and filaments are in redder colors, representing the densest regions of the cosmic web, while those in lower-density regions (i.e., voids) are in blue.

\subsubsection{Tidal environment}
\label{sec:tidal}
The tidal field describes the gravitational pull exerted by the global distribution of matter around a point. The tidal tensor around a subhalo can be mathematically defined as \citep[e.g.,][]{Paranjape2018, Martizzi2019}:
\begin{equation}
T_{ij}(\vec{x})=\partial_{i}\partial_{j}\psi_R(\vec{x}),
\label{eq:tidal_tensor}
\end{equation}
where $\psi_R(\vec{x})$ is the normalised gravitational potential smoothed at a scale $R$ (in what follows we assume a Gaussian smoothing). In order to evaluate the tidal tensor, we need to invert the Poisson equation
\begin{equation}
\nabla^2 \psi_R(\vec{x})=\delta_R(\vec{x}),
\label{eq:poisson}
\end{equation}
where $\delta_R(\vec{x})$ is the smoothed density contrast that in Fourier becomes:
\begin{equation}
\delta_R(\vec{k})=\delta(\vec{k})e^{-k^2R^2/2}.
\label{eq:densityFourier}
\end{equation}
In terms of the Fourier quantities above, Eq.\,\ref{eq:tidal_tensor} becomes:
\begin{equation}
T_{ij}(\vec{x})={\rm FFT}\left[(k_ik_j/k^2)\delta(\vec{k}) e^{-k^2R^2/2}\right],
\label{eq:tidal_tensor_Fourier}
\end{equation}
where $e^{-k^2R^2/2}$ is the Gaussian smoothing filter and $\delta(\vec{k})$ is the Fourier transform of the real-space density field $\delta(\vec{x})$ that we interpolate on a cubic grid with $N_{\rm g}=1024^3$ cells, using a clouds-in-cells (CIC) approach. For this specific task we use the \textsc{Spider} code developed by \citet{Martizzi2019} and publicly available on GitHub\footnote{\url{https://github.com/dmartizzi/spider-public}}. 

\begin{figure}
    \centering\vspace{-0.2cm}
        \includegraphics[width=0.96\linewidth]{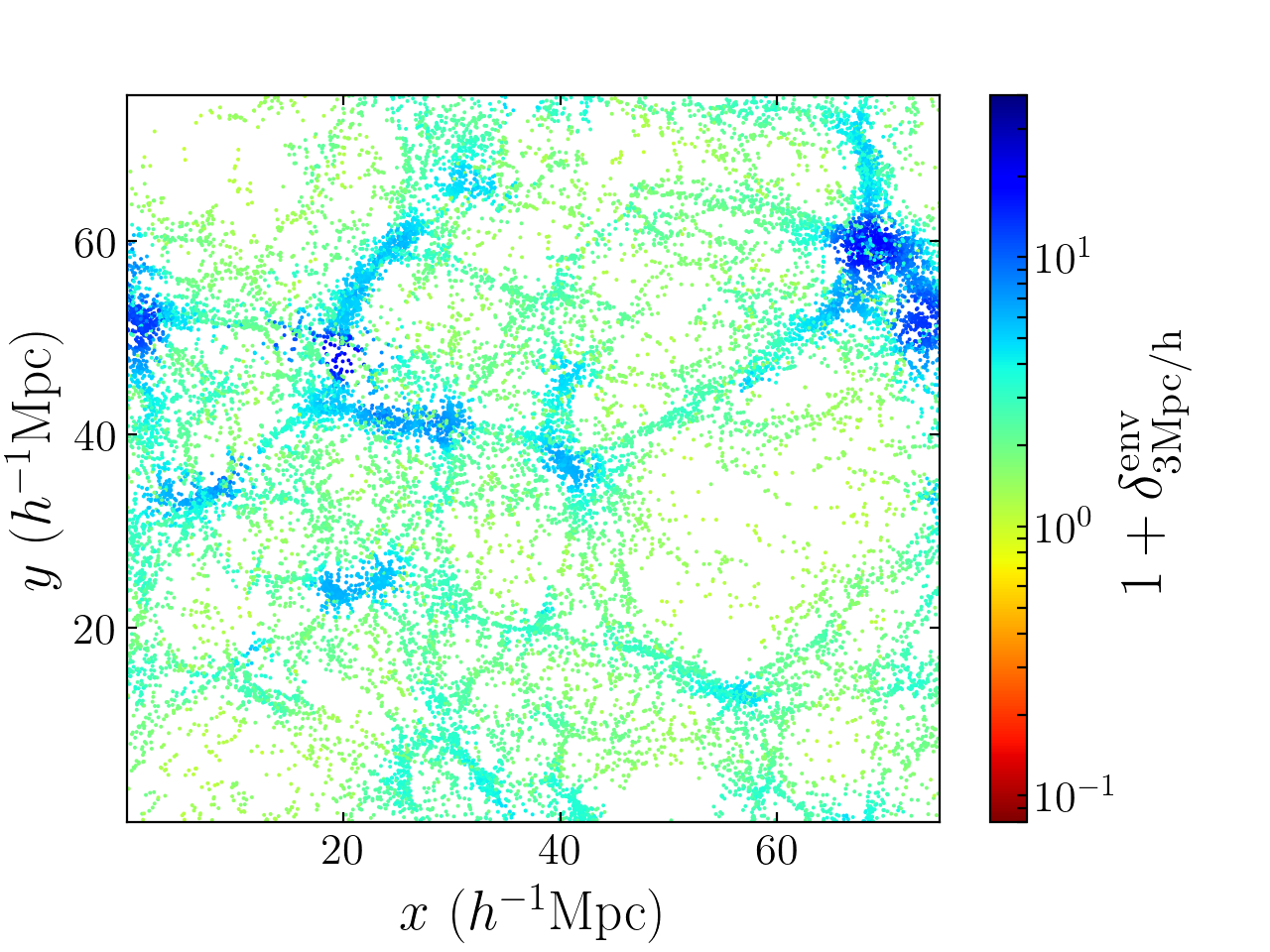}\vspace{-0.1cm}
        \includegraphics[width=0.96\linewidth]{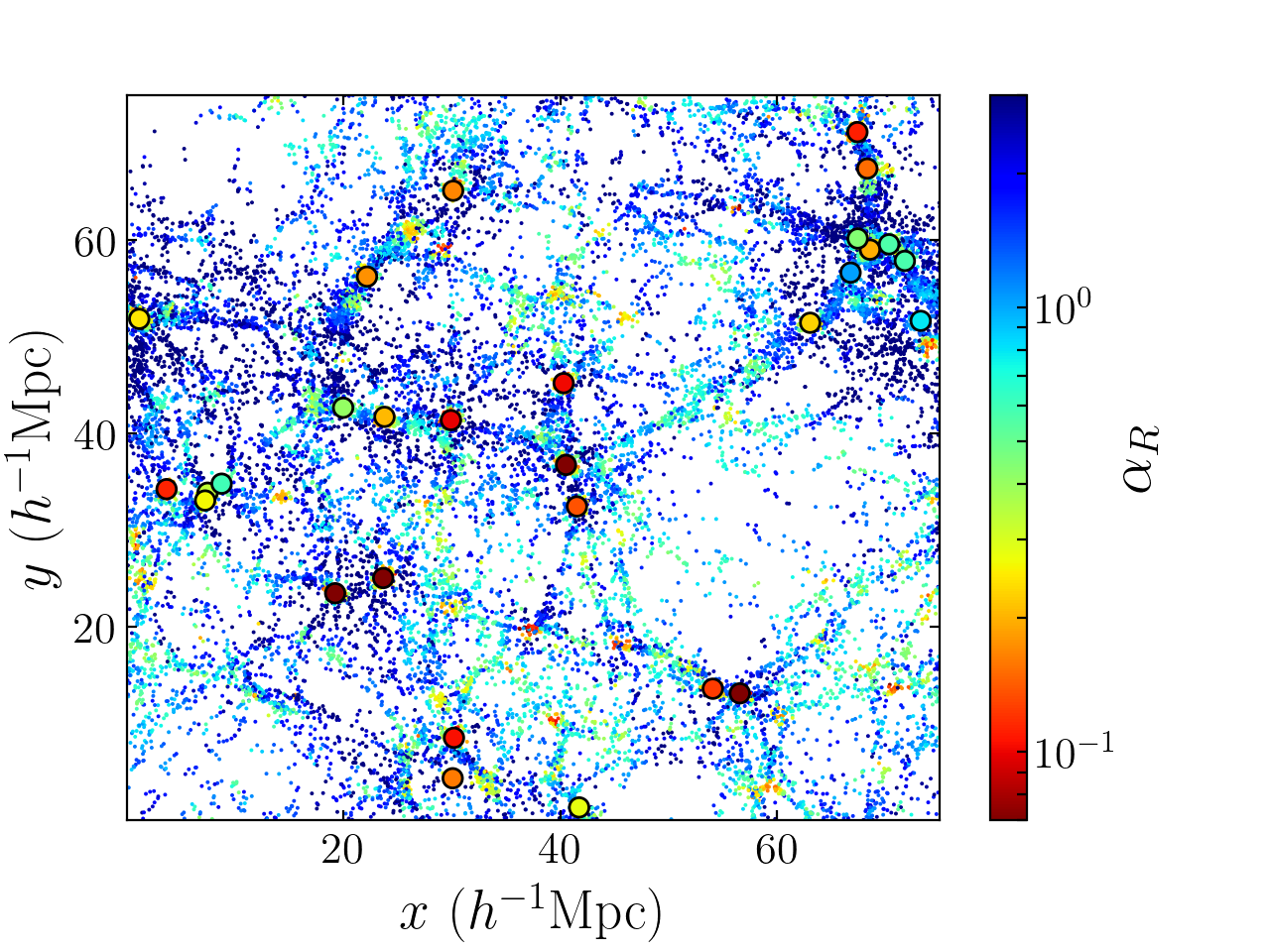}\vspace{-0.1cm}
            \includegraphics[width=0.96\linewidth]{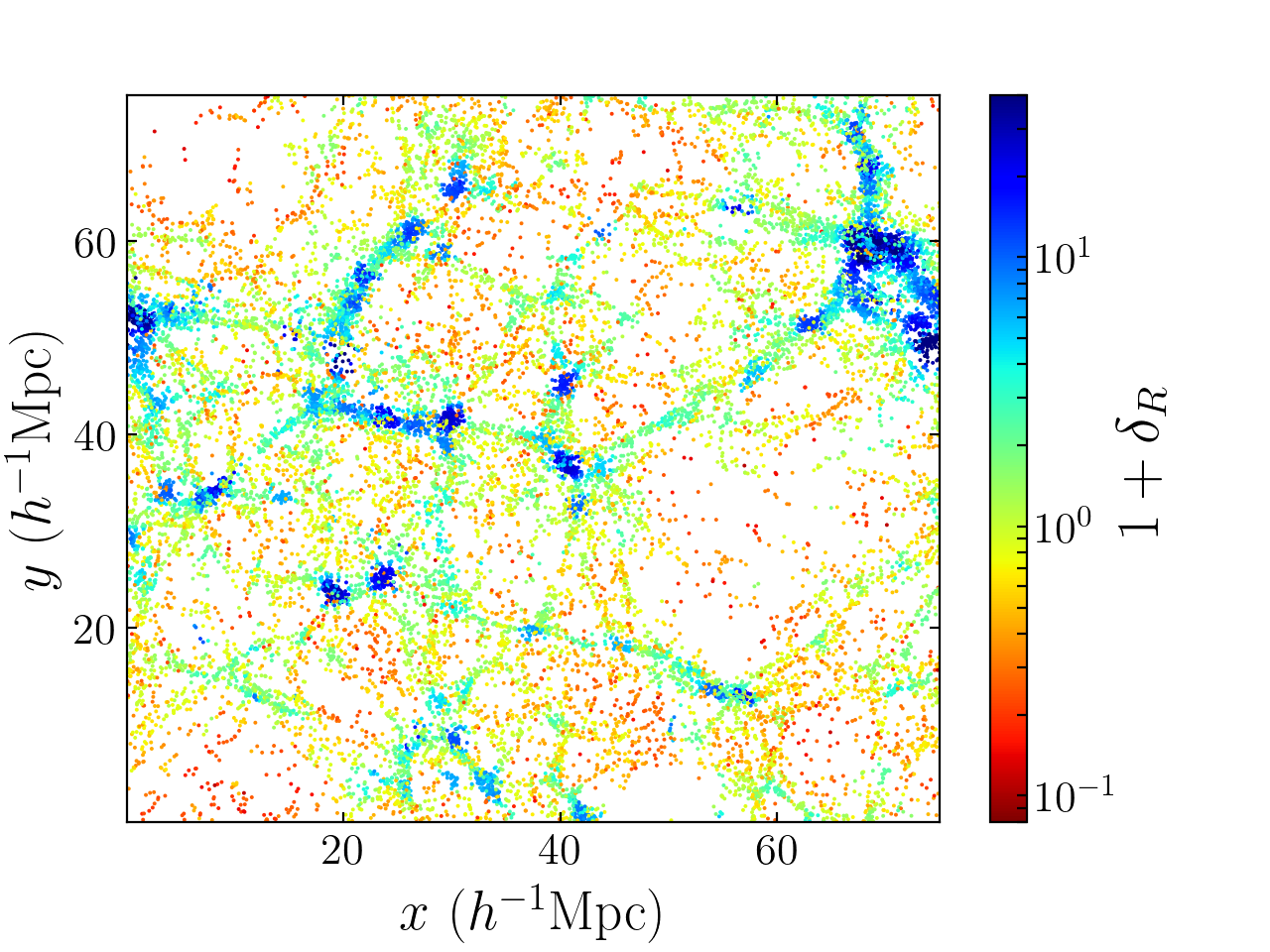}
        \caption{A slice of TNG100, 10$h^{-1}$Mpc thick, color-coded using the environment overdensity at 3$h^{-1}$Mpc (top), the tidal anisotropy parameter $\alpha_R$ (centre), and the tidal over density $\delta_R$ (bottom). We show only the low-mass subhaloes, i.e. M$_{\rm sh}<10^{12}\,h^{-1}$M$_{\odot}$. The big points in the bottom panel indicate the most massive (M$_{\rm sh}>10^{13}\,h^{-1}$M$_{\odot}$) knots.}
        \label{fig:ad}
\end{figure}

Following \citet{Paranjape2018}, \citet{Ramakrishnan2019} and \cite{Zjupa2020}, we smooth the gravitational potential using a range of 15 fixed Gaussian filters log-spaced in $2\,h^{-1}{\rm Mpc}\le R\le 5\,h^{-1}{\rm Mpc}$. The minimum value of this range is safely above the lower resolution limit of the TNG100 subhalo size, $R_{\rm res}=L_{\rm box}/N_{\rm g}^{1/3}=73h^{-1}$kpc,  and was chosen to maximise the contrast between filaments and knots in the $\alpha_R$ map shown in the middle panel of Fig.\,\ref{fig:ad}. We then interpolate the potential in configuration space at each subhalo location ($x_{\rm sh}$, $y_{\rm sh}$, $z_{\rm sh}$) and the smoothing scale at each subhalo radius, $R_{\rm sh}$, to create a subhalo-by-subhalo catalogue of tidal tensor estimates. 

For the $R_{\rm sh}$ interpolation we adopt the Gaussian equivalent of $4R_{\rm 200b}$ defined in terms of the TNG100-DMO subhalo mass as \citep{Paranjape2018}:
\begin{equation}
\begin{aligned}
    R_{\rm G,eff}^{\rm (4R_{200b})}&=\frac{4R_{\rm 200b}}{\sqrt{5}}=\\
    &=1212\,[h^{-1}{\rm kpc}]\left( \frac{M_{\rm sh}}{ 2\times10^{13}[h^{-1}{\rm M_{\odot}}]}\right)^{1/3}\left( \frac{ 0.276}{\Omega_m}\right)^{1/3}.
    \label{eq:Reqv}
    \end{aligned}
\end{equation}
where $\Omega_m$ is the IllustrisTNG100 matter density value and the mass of a resolved subhalo can be written as a function of the number of DM particles that compose it, $N_{\rm p}^{\rm{(halo)}}$, as:
\begin{equation}
\begin{aligned}
    M_{\rm sh}=3.8524\times 10^{11}[h^{-1}M_{\odot}]&\left(\frac{N_{\rm p}^{\rm{(halo)}}}{200}\right)\left(\frac{1024^3}{N_{\rm p}}\right)\times\\
    &\times\left(\frac{\Omega_m}{0.276}\right)\left(\frac{L_{\rm box}}{300h^{-1}{\rm Mpc}}\right)^3.
    \end{aligned}
\end{equation}
In the expression above, $L_{\rm box}$ is the TNG100 box length and $N_{\rm p}$ its total number of DM particles (see Sec.\,\ref{sec:data}).

The cubic lattice applied on the TNG100 volume defines $1024^3$ grid cells. The number of grid cells enclosed in a sphere of radius $2R_{\rm 200b}$ can be written as \citep[see][]{Paranjape2018}:
\begin{equation}
N_{\rm encl}(R_{\rm 200b})=\left(\frac{N_{\rm p}^{\rm{(halo)}}}{200}\right)\left(\frac{1024^3}{N_{\rm p}}\right)\left(\frac{N_{\rm g}}{512^3}\right).
\end{equation}
In line with \citet{Paranjape2018} and \citet{Ramakrishnan2019}, we maximise the correlation between the subhalo tidal properties and the large-scale bias by requiring that $N_{\rm encl}(R_{\rm 200b})\geq 8$ for a TNG100-DMO subhalo to be resolved. This cut, which is equivalent to $N_{\rm p}^{\rm{(halo)}}\geq 1123$ and $R_{\rm G,eff}^{\rm (4R_{200b})}\geq 81h^{-1}$\,kpc, removes about $24\%$ of the 230136 TNG100-DMO subhaloes (48817 centrals; 181319 satellites) surviving the minimal cuts in Sec.\,\ref{sec:selection}. Of these excluded subhaloes, $\sim88\%$ (47939) are satellites and $\sim12\%$ (6719) are centrals.

We then diagonalize the subhalo-centric tidal tensor to extract its eigenvalues $\lambda_1 \leq \lambda_2 \leq \lambda_3$, which give us the following classification of the subhalo environment at scale $R_{\rm sh}$ \citep[see][]{Martizzi2019}:
\begin{equation}
    \begin{aligned}
    &{\rm knots:}\,\,\,\lambda_{\rm i,j,k} \ge \lambda_{\rm th}\\
        &{\rm filaments:}\,\,\,\lambda_{\rm i,j}\ge \lambda_{\rm th}\\
&{\rm sheets:}\,\,\,\lambda_{\rm i}\ge \lambda_{\rm th}\\    
&{\rm voids:}\,\,\,\lambda_{\rm i,j,k}< \lambda_{\rm th},
    \end{aligned}
\end{equation}
where the threshold $\lambda_{\rm th}$ is a free parameter that needs to be adjusted for different smoothing scales. As \citet{Martizzi2019} and \citet{Forero-Romero2009}, we set a fiducial value of $\lambda_{\rm th}=0.3$.

Finally, we use the tidal eigenvalues above to define the subhalo-centric overdensity $\delta_R$ as \citep{Paranjape2018}:
\begin{equation}
    \delta_R=\lambda_1+\lambda_2+\lambda_3,
    \label{eq:eq_deltaR}
\end{equation}
and the tidal shear $q^2$ as:
\begin{equation}
    q^2=\frac{1}{2}\left[(\lambda_2-\lambda_1)^2+(\lambda_3-\lambda_1)^2+(\lambda_3-\lambda_2)^2 \right].
    \label{eq:eq_q2}
\end{equation}
From these two quantities we infer the tidal anisotropy parameter $\alpha_R$ as \citep{Paranjape2018,Ramakrishnan2019}: 
\begin{equation}
    \alpha_R=\sqrt{q^2}/(1+\delta_R).
    \label{eq:eq_alphaR}
\end{equation}

In the middle panel of Figure\,\ref{fig:ad}, we present a slice of the TNG100-DMO simulation, 10$h^{-1}$Mpc thick, color-coded by the anisotropy parameter of the subhalo tidal field, $\alpha_R$. Here we show only the lower-mass subhaloes, i.e. M$_{\rm sh}<10^{12}\,h^{-1}$M$_{\odot}$. In typically bluer colours are the filaments and sheets characterised by higher anisotropy values, while in redder are the knots, the densest and most isotropic regions of the large-scale structure. We overplot as big points the most massive (i.e., M$_{\rm sh}>10^{13}\,h^{-1}$M$_{\odot}$) haloes. Despite the differences in the simulation, grid and sample selection, our result is overall consistent with Fig.\,9 in \citet{Paranjape2018}. The biggest discrepancy is that, in our case, the field is characterised by higher anisotropy values, which might be explained by the different configuration, resolution and range of smoothing radii adopted.

In the bottom plot we show the same map color-coded using the tidal overdensity, $\delta_R$, which traces particularly well the high-density regions around the knots. Blue and light-blue colours mark the higher-density filaments and sheets, whereas galaxies in the field are, as expected, associated with low values of $\delta_R$.

\section{Methodology}
\label{sec:methodology}
\subsection{Galaxy clustering measurements} 
\label{sec:clustering_measurements}
We measure the real-space two-point correlation function (2PCF) of the TNG100 galaxies using the code implemented by \citet{Favole2016cameron}. This is based on the natural estimator \citep{1974ApJS...28...19P}:
\begin{equation}
    \xi(r)=\frac{DD(r)}{RR(r)}-1,
\end{equation}
where $DD$ and $RR$ are the normalised data-data and random-random pair counts, respectively. Taking advantage of the cubic geometry of the simulation, we approximate $RR$ using the spherical shells as \citep[e.g.,][]{1986ApJ...301...70R}:
\begin{equation}
    RR(r)=\frac{{\rm dVol}(r)}{{\rm Vol}}=\frac{4\pi}{3}\frac{[(r+dr)^3-r^3]}{ L_{\rm box}^3},
\end{equation}
where $L_{\rm box}=75\,h^{-1}$Mpc is the side dimension of the IllustrisTNG100 simulation box.

We estimate the uncertainties on the galaxy clustering measurements via jackknife resampling following the procedure adopted by \citet{Hadzhiyska2020}. We divide the TNG100 volume in $3^3=27$ sub-boxes with size $(75/3)\,h^{-1}\rm Mpc/\approx25\,{\textit h^{-1}}\rm Mpc$. We then compute the 2PCF of the set of cubes eliminating a different one each time. We obtain the uncertainty on the 2PCF of the full TNG100 box from the diagonal elements of its covariance matrix, defined as:
\begin{equation}
    C_{ij}(r)=\frac{\rm N_{\rm res-1}}{\rm N_{\rm res}}\sum_{a=1}^{\rm N_{\rm res}}[\xi_i^a(r)-\bar{\xi_i}(r)][\xi_j^a(r)-\bar{\xi_j}(r)],
\end{equation}
where $\rm N_{\rm res}=27$ and $\bar{\xi_i}(r)$ is the mean jackknife 2PCF in the $i^{\rm th}$ spatial bin:
\begin{equation}
    \bar{\xi_i}(r)=\frac{1}{\rm N_{\rm res}}\sum_{a=1}^{\rm N_{\rm res}}\xi_i^a(r).
\end{equation}


\subsection{Multi-population SHAM}
\label{sec:multiSHAM}
\begin{figure*}
\begin{center}
\includegraphics[width=0.96\linewidth]{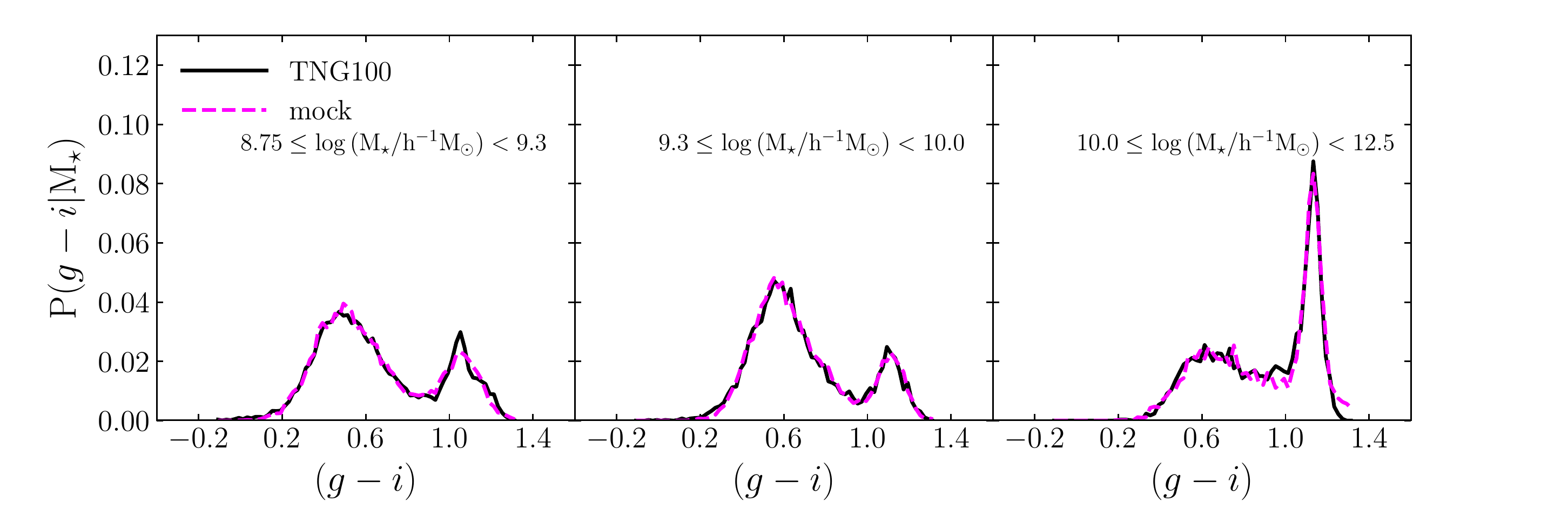}
\includegraphics[width=0.96\linewidth]{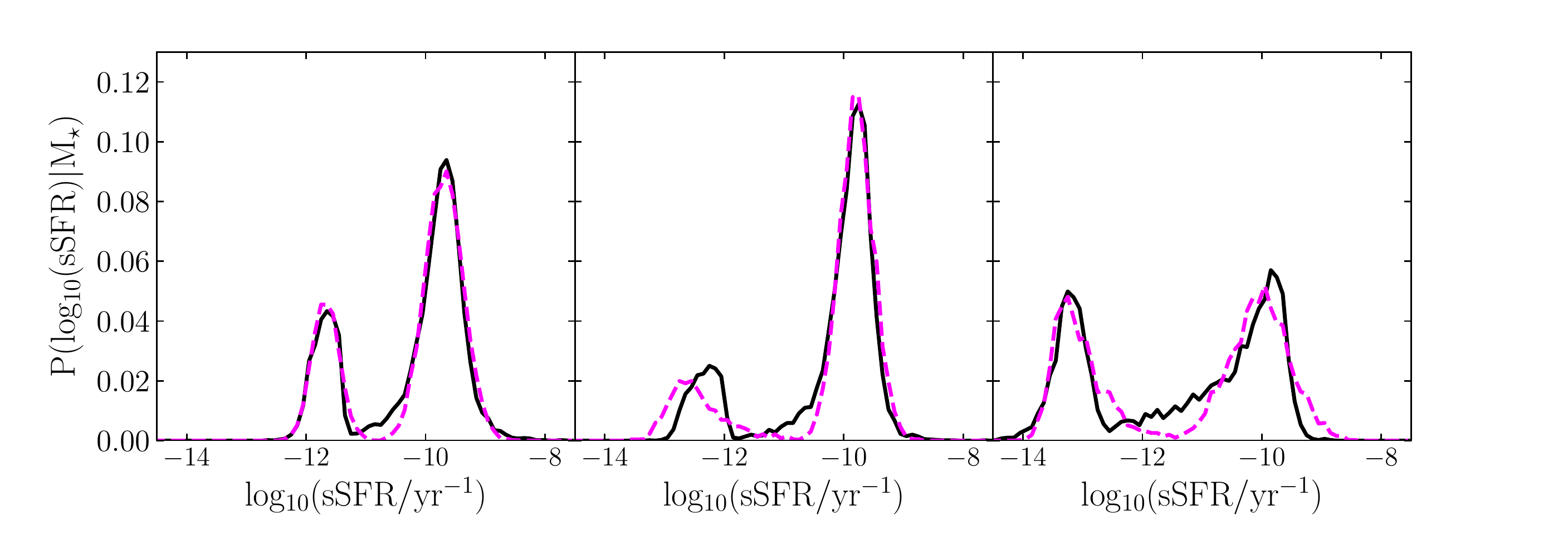}
\caption{Conditional probability distribution of the TNG100 galaxies (solid black lines) as a function of the secondary galactic properties in bins of stellar mass. The dashed magenta curves are the draws for the mocks. From top to bottom we show the PDFs for the galaxy color and sSFR.}
\label{fig:pdfs}
\end{center}
\end{figure*}
The sub-halo abundance matching is a straightforward prescription used to connect galaxies with their host dark-matter subhaloes based on the simple assumption that more massive (or luminous) galaxies reside in more massive subhaloes. The standard SHAM assignment is performed by rank-ordering galaxies and subhaloes according to specific primary properties and by matching their cumulative number densities. In this analysis, we adopt as properties the galaxy stellar mass and the subhalo maximum circular velocity over its entire history, $V_{\rm{peak}}$, which has been shown to perform better than other subhalo proxies \citep{Contreras2020,Hadzhiyska2020}.

In order to ensure that our model is physically plausible, we allow for a constant Gaussian scatter $\sigma_{V}$ in the $M_{\star}-V_{\rm peak}$ relation. In practice, the procedure consists in randomly sampling galaxies from the TNG100 cumulative stellar mass function (see Fig.\,\ref{fig:smf}) convolved with a Gaussian probability distribution function (PDF) with fixed amplitude $\sigma=0.125$ dex. This specific value was chosen to match the results obtained by \cite{Contreras2020}.

The aforementioned standard SHAM is extended in order to incorporate secondary subhalo and galaxy properties, following a similar methodology to that laid down by \cite{Hearin2013}. We call this methodology \textit{multi-population SHAM}. First, we divide the TNG100 galaxies in three bins of stellar mass designed to have high-enough number density:
$8.75\leq {\rm log(M_{\star}/h^{-1}M_{\odot})<9.3}$, $9.3\leq {\rm log(M_{\star}/h^{-1}M_{\odot})<10.0}$ and $10.0\leq {\rm log(M_{\star}/h^{-1}M_{\odot})<12.5}$. In each bin, we define the conditional PDF for a galaxy with a given stellar mass to have a specific secondary property, $P(X|\rm M_{\star})$, where $X$ is the secondary property considered. The secondary properties we explore are $(g-i)$ colour and sSFR. 

Figure \ref{fig:pdfs} presents the TNG100 PDFs (solid black lines) in bins of stellar mass for each one of the galaxy secondary properties. Both the colour and sSFR distributions exhibit a clear bimodality with a red (quenched) and a blue (star-forming) peak. We will use these trends in the analysis to split the full sample into two populations with different colour/sSFR and to study the dependence of galaxy clustering on such properties.

The question is whether the secondary galaxy dependence introduced above can be accounted for by a secondary subhalo property. In order to test this, we implement the secondary matching through the following steps:
\begin{enumerate}
    \item fit the TNG100 PDFs using composite Gaussian functions and derive the analytic PDFs.\\
    \item split the mock catalogue obtained from the basic SHAM in the three stellar-mass bins defined above and, in each bin, rank-order the mocks according to the secondary subhalo property we want to match. These are: z$_{\rm starve}$, c$_{\rm infall}$, $\delta_R^{\rm env}$, $\alpha_R$, and $\delta_R$.\\
    \item for each mock galaxy, we draw a secondary galaxy property from the analytic PDFs defined above.\\
    \item rank-order the draws (dashed magenta lines in Fig. \ref{fig:pdfs}) and assign them to the mocks. In this way the correlation between galaxy and subhalo secondary properties at fixed stellar mass is preserved. 
\end{enumerate}


\section{Results}
\label{sec:results}

\subsection{Correlations between secondary properties}
\label{sec:secpropcorrel}
\begin{figure*}
\centering
\includegraphics[width=\linewidth]{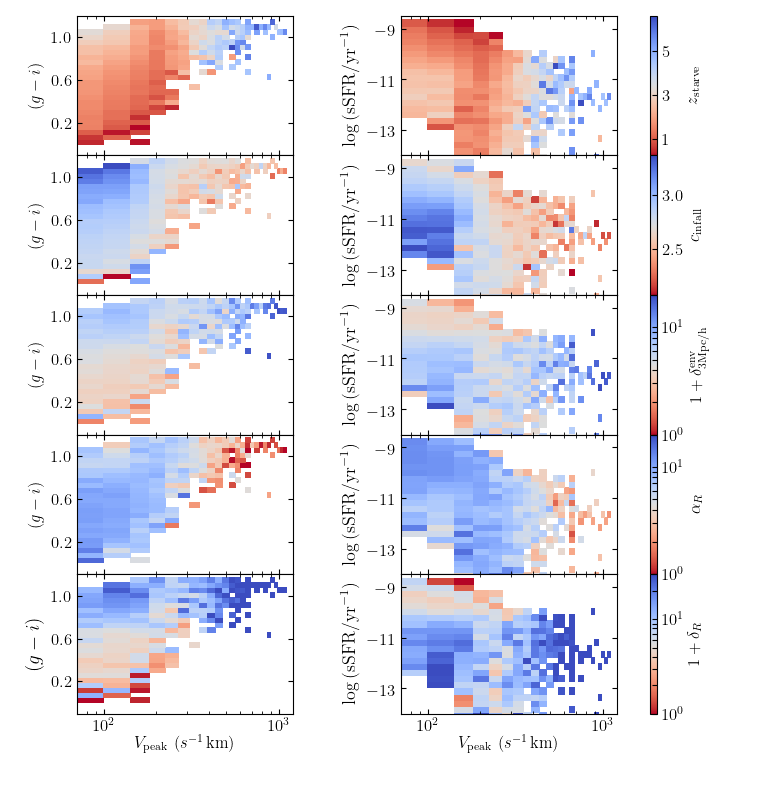}\vspace{-1cm}
\caption{Secondary properties of the IllustrisTNG100 galaxies (i.e. colour and sSFR) as a function of $V_{\rm peak}$, color-coded with the secondary halo properties (i.e., z$_{\rm starve}$, c$_{\rm infall}$, $\delta_{\rm 3Mpc/h}^{\rm env}$, $\alpha_R$ and $\delta_R$). For the tidal properties we consider only subhaloes above the resolution limit $R_{\rm G,eff}^{\rm (4R_{200b})}\geq 81h^{-1}$\,kpc (see Sec.\,\ref{sec:tidal}). Here we show the average value of the secondary halo property in 30 bins of $V_{\rm peak}$ and the secondary galaxy property.}
\label{fig:secondprops}
\end{figure*}
We study how the IllustrisTNG100 galaxy and subhalo secondary properties correlate as a function of $V_{\rm peak}$. This quantity is fundamental for our analysis, as it will be used as main proxy for the subhaloes in the SHAM assignment (Sec. \ref{sec:multiSHAM}).

Figure \ref{fig:secondprops} summarises our findings, together with Table\,\ref{tab:correlcoeff}, where we report the correlation coefficients at fixed $V_{\rm peak}$. We observe strong correlation between the galaxy colour $(g-i)$ and the subhalo z$_{\rm starve}$, with redder galaxies undergoing starvation at higher redshift. A weaker correlation is observed between the specific SFR (i.e. SFR per unit stellar mass) and z$_{\rm starve}$. Here we see that starvation happens at lower (higher) redshift for star-forming (quenched) galaxies. 

Good correlation is observed also between the galaxy colour and the subhalo concentration at infall, with reddest galaxies in the low-$V_{\rm peak}$ end showing the highest concentrations. A similar but weaker trend is also found between sSFR and concentration in the low-$V_{\rm peak}$ regime.

A similar correlation, lower than in the previous cases, is appreciable between the galaxy color and both halo overdensities $\delta_{\rm 3Mpc/h}^{\rm env}$ and $\delta_R$, confirming that redder galaxies preferentially occupy densest regions of the cosmic web. This trend is progressively washed out as we increase the radius $R$ at which $\delta_R^{\rm env}$ is computed. A weaker correlation is observed between sSFR and both subhalo overdensities.

Between the explored subhalo properties, the tidal anisotropy $\alpha_R$ is the one that correlates the least with both the galaxy color and the sSFR. Some trend is visible, but weaker compared to the rest of subhalo proxies. Note that due to the stochasticity in the galaxy formation process and in the connection between haloes and their LSS environments, a very high level of correlation between the properties under analysis is, of course, not expected. We are, however, looking for signs that could eventually lead to further refinements of the SHAM procedure.

Our results tell us that, overall, there is a tendency for quenched, redder galaxies to inhabit the denser regions in the cosmic web, which are usually more isotropic.
The correlations observed between the IllustrisTNG100 galaxy and subhalo secondary properties, even if mild, will play a crucial role in our SHAM modelling, as they will be used as drivers to correctly draw the secondary properties of the mock galaxies.

\begin{figure*}
    \centering
    \includegraphics[width=\linewidth]{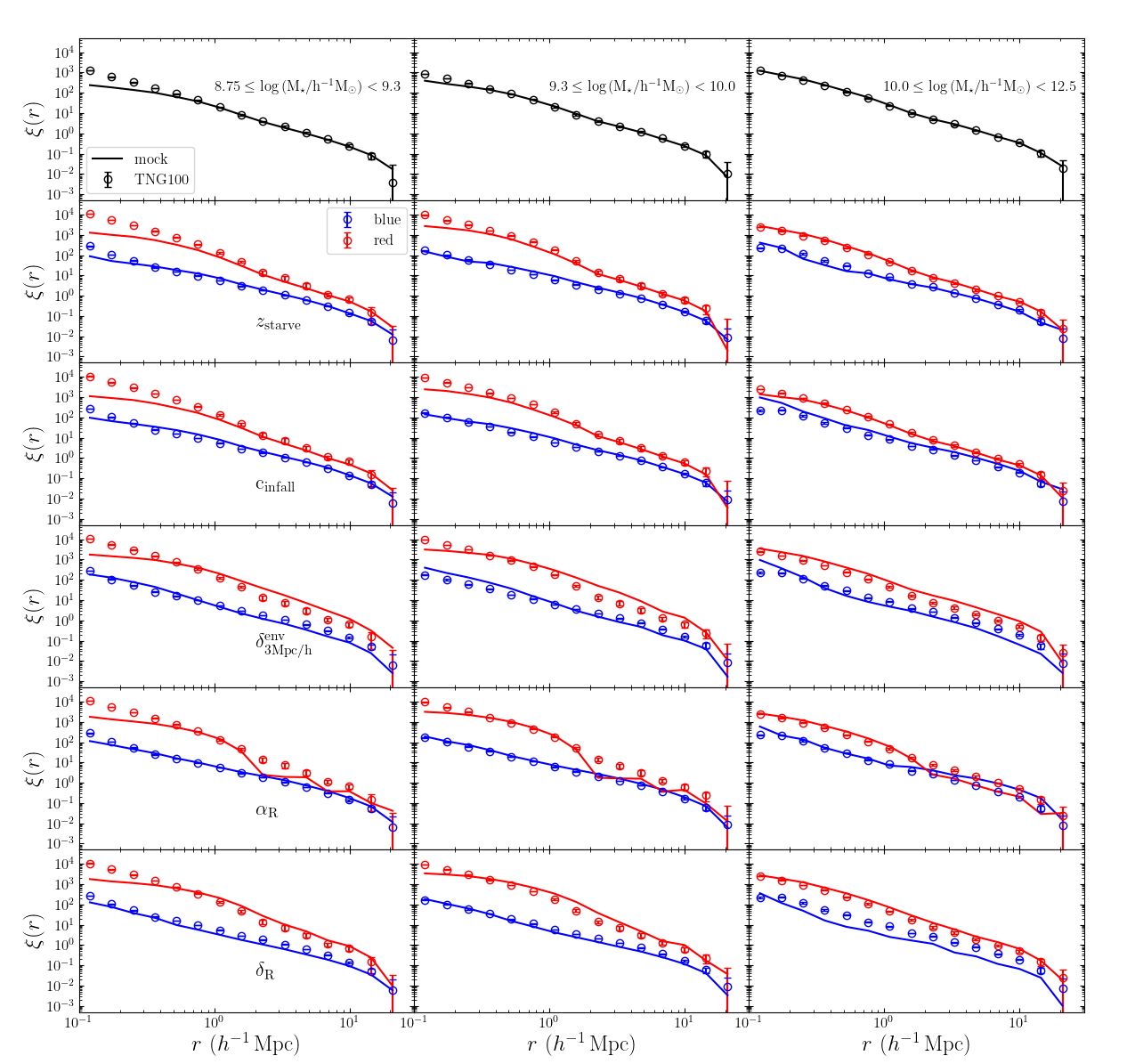}\vspace{-0.5cm}
        \caption{Two-point correlation functions of the IllustrisTNG100 galaxies (points) and mocks (lines) in three stellar mass bins (columns). In the top row we show the full galaxy population in each mass bin against the mock resulting from basic SHAM. In the other panels we show the results for the red/blue galaxy samples and the SHAM including secondary matching performed between the galaxy colour and the halo properties indicated in the first panel of each row. From top to bottom we show: z$_{\rm starve}$, c$_{\rm infall}$, $\delta_{\rm 3Mpc/h}^{\rm env}$, $\alpha_R$, $\delta_R$. The shaded error bars are estimated performing 27 jackknife resamplings in IllustrisTNG100 data.}
    \label{fig:primo}
\end{figure*}
\begin{figure*}
    \centering
    \includegraphics[width=\linewidth]{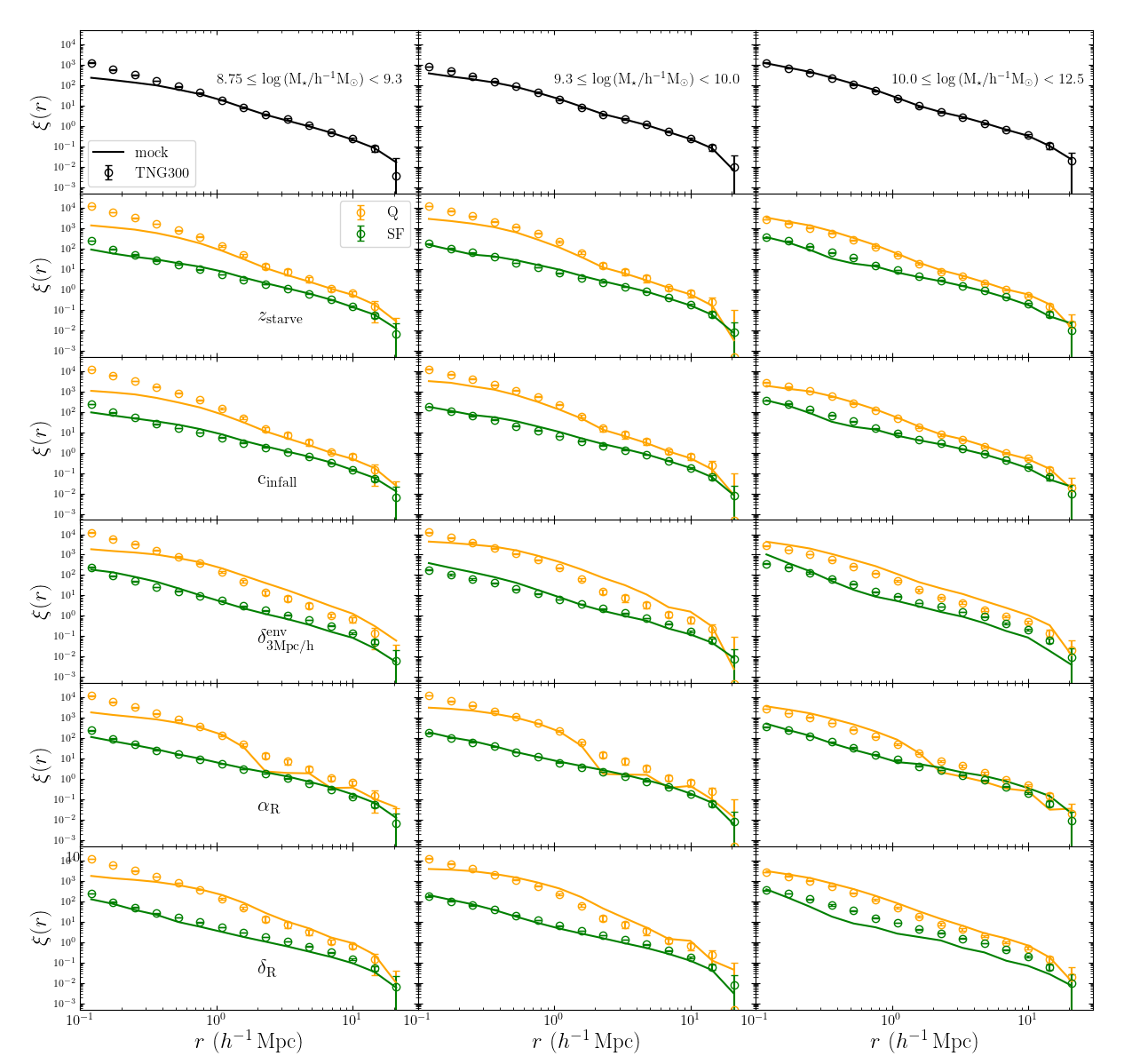}\vspace{-0.5cm}
        \caption{Same results as in Fig.\,\ref{fig:primo}, but with sSFR as secondary galaxy property.}
        \label{fig:secondo}
\end{figure*}

\begin{table}\centering
\begin{tabular}{@{}lcc@{}}
\toprule
& $(g-i)$&$\log{\rm (sSFR/yr^{-1})}$ \\
\midrule
$z_{\rm starve}$&0.56&-0.36\\
$c_{\rm infall}$&0.49&-0.30\\
$1+\delta_{\rm 3Mpc/h}^{\rm env}$&0.46&-0.26\\
$\alpha_R$&-0.22&0.11\\
$1+\delta_R$&0.46&-0.25\\
\bottomrule
\end{tabular}
\caption{Correlation coefficients between the secondary subhalo and galaxy properties at fixed $V_{\rm peak}$ in the range 100-200$\,s^{-1}$km, where 68\% of the subhaloes lie.}
\label{tab:correlcoeff}
\end{table} 

\subsection{Galaxy clustering results}
\label{sec:clustering_results}
In what follows we present the two-point correlation functions of the TNG100 galaxies in bins of stellar mass modelled using the secondary matching explained in Sec.\,\ref{sec:multiSHAM} on top of a standard SHAM based on M$_{\star}$ and $V_{\rm peak}$. 

Fig.\,\ref{fig:primo} presents the 2PCF results with secondary matching in colour. In the top row we show the clustering of the full TNG100 galaxy sample (points) modelled using standard SHAM (lines) and, in the rest of panels, the results for the sub-samples separated in colour and modelled with a secondary matching with the following subhalo properties: starvation redshift z$_{\rm starve}$, concentration at infall c$_{\rm infall}$, subhalo overdensity $\delta_{\rm 3Mpc/h}^{\rm env}$, tidal anisotropy $\alpha_R$ and tidal overdensity $\delta_R$. 

Even if apparently mild, the correlations observed in Fig.\,\ref{fig:secondprops}, between galaxy and subhalo secondary properties, return a good agreement in the clustering amplitude of the red/blue TNG100 populations and our mock catalogues over the entire stellar mass range. In all the cases explored we are able to recover a clear separation in the amplitude of the red and blue models, consistent with the TNG100 fiducial data sets.

It is noteworthy that in the lowest $M_{\star}$ bin all the full, red and blue models underestimate the TNG100 small-scale ($r\lesssim0.2\,h^{-1}$Mpc) clustering amplitude, due to a lack of satellite subhaloes. This effect is also present, even if reduced, in the intermediate mass bin, in particular in the full and red mock catalogues.

Those subhalo secondary properties more tightly correlated with the galaxy color and sSFR, e.g. $z_{\rm starve}$, $c_{\rm infall}$ and $\delta_R$, perform better in reproducing the 2PCF separation. The tidal anisotropy $\alpha_R$ behaves very well in reproducing the clustering of the blue population, which is the densest one, in all three mass bins. The red model performs less well, highlighting the transition between the 1- and 2-halo term with a pronounced bump and, on larger scales, its fluctuations denote a lack of massive objects. Overall, both subhalo tidal properties demonstrate to be valuable tracers of the large-scale structure. Interestingly, the SHAM correlation functions cross over on large scales for the highest mass bin, where bluer objects are predicted to be more tightly clustered than redder objects. This intriguing result will be investigated in more depth in follow-up work.

It is not surprising that the clustering models based on $\delta_R^{\rm env}$ are similar but not identical to those based on $\delta_R$, as the two subhalo overdensities are defined in different ways. In fact, while $\delta_R^{\rm env}$ is obtained by counting subhaloes within a fixed smoothing scale (we tested the values $R=3,5,8\,h^{-1}$Mpc), the tidal quantity $\delta_R$ is inferred by diagonalising the subhalo tidal tensor interpolated at the subhalo positions, and smoothed at the Gaussian equivalent of 4$R_{\rm 200}$ for each subhalo (see \S\,\ref{sec:tidal}).

Fig.\,\ref{fig:secondo} presents the TNG100 clustering results as a function of sSFR. Also in this case, the subhalo secondary properties that correlate the most with sSFR, and hence return the best clustering models, are z$_{\rm starve}$, {\bf $c_{\rm infall}$} and $\delta_R$. Overall the subhalo properties perform equally well when coupled to galaxy colour or sSFR.


\section{Summary}
\label{sec:discussion}
In order to model the clustering of multiple populations of galaxies, the standard SHAM prescription needs to be modified to account for the different distributions of their galaxy properties at fixed stellar mass. In this work, we use the IllustrisTNG100 hydrodynamical simulation to test the SHAM performance in bins of stellar mass as a function of different galaxy and subhalo properties. 

First, we have implemented a standard SHAM on the TNG100 galaxy population split in three stellar mass bins using M$_{\star}$ and $V_{\rm peak}$ as primary proxies for galaxies and subhaloes, respectively. We have chosen $V_{\rm peak}$, that is, the maximum circular velocity over the entire history of the subhalo, because it is proven  
to perform better as a subhalo proxy in SHAM compared to the halo maximum circular velocity or the infall velocity \citep{Chaves-Montero2016,Hadzhiyska2020}. 
 
 In each stellar mass bin, we have divided the TNG100 galaxies into two colour (red/blue) and sSFR (quenched/star-forming) sub-samples. For these sub-samples, we have measured the clustering and modelled the results by implementing a decorated SHAM assignment capable of coupling secondary galactic and subhalo properties. This secondary matching is an extension of the {\it{age matching}} prescription by \citet{Hearin2013}, but including several other secondary properties: galaxy sSFR, halo c$_{\rm infall}$, $\delta_{3,5,8{\rm Mpc/h}}^{\rm env}$, $\alpha_R$ and $\delta_R$. 
 
 As a result, we have overall found good agreement between our mocks and the TNG100 observations. In particular, the accuracy of the models depends on the galaxy and subhalo secondary properties adopted for the matching. We summarise our findings as follows:
 
  \begin{itemize}
  \item Among the secondary subhalo properties studied, at fixed stellar mass, we find that the starvation redshift z$_{\rm starve}$ and the the concentration at infall $c_{\rm infall}$ qualitatively provide the best clustering results. Our z$_{\rm starve}$ outcome confirms previous results from \citet{Hearin13}.
  \item Other physically motivated subhalo properties, such as the tidal overdensity $\delta_R$ and the subhalo overdensity $\delta_R^{\rm env}$ perform also well, with slightly larger deviations from the fiducial data set. The tidal anisotropy $\alpha_R$ performs well in reproducing the sub-populations with higher number density, while for the lower-density sample the disagreement with TNG100 is more pronounced.
  \item The accuracy of our clustering models obtained through secondary matching improves when the secondary subhalo and galactic properties are tightly correlated.
    \item Although we find interesting signs of correlation, it is still unclear in light of our results how the tidal anisotropy $\alpha_R$ can be efficiently introduced into a SHAM modeling scheme.
     \end{itemize}
The decorated SHAM presented in this work enables robust clustering predictions for different samples of red/blue and quenched/star-forming galaxies. Continuing the development of this methodology is particularly relevant for next-generation surveys, such as DESI\footnote{\url{https://www.desi.lbl.gov}} or Euclid\footnote{\url{https://www.euclid-ec.org}}, which will collect samples of hundreds of millions of galaxies at high redshift. Different galaxy populations act as different tracers of the LSS, hence the importance of adapting our techniques to the new era of {\it{multi-tracer cosmology}}. Our model can be easily extended to match other galaxy/subhalo properties to achieve a more complete vision of the large-scale structure dynamics.

\section*{Acknowledgments}
The authors are thankful to the referee, A. Paranjape, for insightful comments that have been determinant to improve this analysis.

GF acknowledges financial support from the SNF 175751 “Cosmology with 3D Maps of the Universe” research grant. AMD thanks FAPESP and Fondecyt (Fondecyt Regular 2021 grant 1210612) for financial support. MCA acknowledges financial support from the Austrian National Science Foundation through FWF stand-alone grant P31154-N27. SC acknowledges the support of the “Juan de la Cierva Formacion” fellowship (FJCI-2017-33816) and ERC Starting Grant number 716151 (BACCO). IZ acknowledges support by NSF grant AST-1612085.

\section*{Data availability}
 This work made use of the llustrisTNG database available at \url{www.tng-project.org}.
 
\bibliographystyle{mnras}
\bibliography{./references}

\end{document}